\documentclass[fleqn,10pt]{wlscirep}

\usepackage{xcolor}
\usepackage{bm}

\title{Modeling confirmation bias and polarization}
\author[1]{Michela Del Vicario}
\author[1,2]{Antonio Scala}
\author[1]{Guido Caldarelli}
\author[3]{H. Eugene Stanley}
\author[1,*]{Walter Quattrociocchi}
\affil[1]{Laboratory of Computational Social Science, Networks Dept, IMT School for Advanced Studies, 55100 Lucca, Italy}
\affil[2]{ISC-CNR Uos "Sapienza", 00185 Roma, Italy}
\affil[3]{Boston University, Center for Polymer Studies, Department of Physics, Boston, Massachusetts 02215}
\affil[*]{walter.quattrociocchi@gmail.com}

\begin{abstract}
Online users tend to select claims that adhere to their system of beliefs and to ignore dissenting information.
Confirmation bias, indeed, plays a pivotal role in viral phenomena. Furthermore, the wide availability of content on the web fosters the aggregation of likeminded people where debates tend to enforce group polarization.
Such a configuration might alter the public debate and thus the formation of the public opinion.
In this paper we provide a mathematical model to study online social debates and the related polarization dynamics. We assume the basic updating rule of the \textit{Bounded Confidence Model} (BCM) and we develop two variations a) the \textit{Rewire with Bounded Confidence Model} (RBCM), in which discordant links are broken until convergence is reached; and b) the \textit{Unbounded Confidence Model}, under which the interaction among discordant pairs of users is allowed even with a negative feedback, either with the rewiring step (RUCM) or without it (UCM). From numerical simulations we find that the new models (UCM and RUCM), unlike the BCM, are able to explain the coexistence of two stable final opinions, often observed in reality. Lastly, we present a mean field approximation of the newly introduced models.
\end{abstract}

\begin{document}
	
\flushbottom
\maketitle
\thispagestyle{empty}
	
\section*{Introduction}
Online users tend to select claims that adhere to their system of beliefs and to ignore dissenting information \cite{quattrociocchi2016echo,bessi2015science,anagnostopoulos2014viral,zollo2015debunking,josang2011taste}. The wide availability of content on the web fosters the aggregation of likeminded people where the discussion tends to enforce group polarization \cite{zollo2015emotional,sunstein2002law}. Confirmation bias, indeed, plays a pivotal role in viral phenomena \cite{del2015spreading}. 
Under such conditions public debates, in particular on social relevant issues, tend to further fragment and polarize the public opinion \cite{konig2009effects,paolucci2009social}.
	
To better understand this process, in this paper we provide a mathematical model mimicking polarization in online social dynamics.
	
Opinion dynamics, have been widely investigated in recent years, using different approaches from statistical physics and network science \cite{castellano2009statisctical}. Classical examples of opinion dynamics models include the Sznajd model \cite{sznajd2000opinion}, the voter model \cite{holley1975ergodic, ligget1997stochastic, lambiotte2008dynamics}, the majority rule model \cite{krapivsky2003dynamics,galam2008sociophysics}, and the bounded confidence model (BCM) \cite{deffuant2000mixing, hegselmann2002opinion, lorenz2007continuous}. Besides the different assumptions and dynamics rules, for all the cited models the consensus state, in which all agents share the same opinion, is reached under the right conditions.
	
However, consensus in far from common in real world and Internet based opinion exchanges. 
A recent study showed the emergence of \textit{polarized communities}, i.e., \textit{echo chambers}, in online social networks \cite{del2015spreading}. Inside these communities, homogeneity appears to be the primary driver for the diffusion of contents. Both polarization and homogeneity might be the result of the conjugate effect of \textit{confirmation bias} and \textit{social influence}. Confirmation bias is the tendency to acquire or process new information in a way that confirms one's preconceptions and avoids contradiction with prior belief \cite{nickerson1998confirmation}. Social influence is the process under which one's emotions, opinions, or behaviors are affected by others. In particular, \textit{informational influence} occurs when individuals accept information from others as evidence about reality \cite{centola2010spread, centola2015spontaneous}.
	
Previous studies \cite{shao2009dynamic, li2011strategy} proposed a non consensus opinion model (NCO) that allowed for the stable coexistence of two opinions by also considering the opinion of the user herself when applying the majority rule update \cite{shao2009dynamic}, while in \cite {li2011strategy} the competition between two groups is investigated by the introduction of a set of contrarians in one of the two. The survival of a two-opinions state is studied in \cite{majdandzic2014spontaneous} from a different point of view, considering the emergence of spontaneous recovery of failed nodes and the majority rule update. Both these models assume only two opinion states ($\pm1$) and a majority rule update, with the novelty of accounting for the individual opinion \cite{shao2009dynamic, li2011strategy} and for an external source of influence \cite{majdandzic2014spontaneous}. 
	
For the purpose of this study, we are more interested in capturing the emergence of two (or more) stable opinions out of a wide range of possibilities rather than the predominance of one over another. Hence, we will assume a continuous interval of opinions. Also we are inclined to assume peer-to-peer interactions rather than majority rule as a better proxy of face-to-face and online social interactions.
A model grounded on the BCM and accounting for the interconnection and complexity of the online environment as well as the competition among sources of information is presented in \cite{quattrociocchi2014opinion}.
	
People shape their opinions on the basis of both confirmation bias and social influence, a combination of these two forces generates the observed polarization of communities and homogeneous links \cite{del2015spreading}. Accounting for this phenomenon, we build a model of opinion dynamics and network's evolution that considers both mechanisms and expands itself from the classical \textit{Bounded Confidence Model} (BCM) \cite{deffuant2000mixing}. We consider two variations of the model: the \textit{Rewire with Bounded Confidence Model} (RBCM), in which discordant links are broken until convergence is reached; and the \textit{Unbounded Confidence Model}, under which interaction among discordant pairs of users is allowed and a negative updating rule is introduced, either with the rewiring step (RUCM) or without it (UCM). 
	
The paper is structured as follows. In the first section we provide references to the methods employed and give a brief overview of the BCM and its convergence results. In the \textit{Results and Discussion} section, we first present the new models and give an account of the simulation results, then we present a mean field approximation of the newly introduced models.

\section*{Methods}
\subsection*{Periodic Boundary Conditions}
We consider $N$ agents and a set of initial opinions $x_i, \,i\in\{1,\ldots, N\}$, uniformly distributed in $[0,1]$. If we compare two agents' opinions by the absolute value distance $|x_i - x_j|$, those agents with near boundary opinions will have less concordant peers by definitions. 
We can overcome this problem by using the \textit{Periodic Boundary Conditions} (PBC) and the alternative opinions' distance $|.|_{\tau}:[0,1]\times[0,1]\rightarrow[0,0.5]$ defined as:
$$
	|x_i - x_j|_{\tau} = |x_i - x_j - \rho(x_i -x_j)|,
	$$
for $i,j\in\{1\ldots, N\}$. 
The $\rho(.):[-1,1]\rightarrow\{-1,0,1\}$ adjustment ensures PBC and it is defined as: 
	
	\begin{equation}\label{pbc}
	\rho(x) = 
	\begin{cases}
	-1,\,\,\, \mbox{ if }x\in[-1,-0.5)\\	
	\,\,\,\,	0, \,\,\,\mbox{ if }x\in[-0.5,0.5]\\
	\,\,\,\,	1, \,\,\,\mbox{ if }x\in(0.5,1]\\
	\end{cases}.
	\end{equation}
	
\subsection*{The Bounded Confidence Model (BCM)}
The \textit{Bounded Confidence Model} (BCM) \cite{deffuant2000mixing, lorenz2007continuous} considers a set of $N$ agents arranged on a complex network $G$.\footnote{We consider different types of complex networks in the simulations: The Erd\"{o}s-R\'enyi random network (ER) \cite{erdHos1959random} characterized by a Poisson degree distribution with average degree $\langle 2\rangle$, the Scale-Free network (SF) \cite{barabasi1999emergence} characterized by a power-law degree distribution $P(k)\sim k^{-\gamma}$ with $2\leq\gamma\leq 3$, and the Small-World network (SW) \cite{watts1998collective} with rewiring probability equal to $0.2$ and neighborhood dimension equal to $2$. We notice that the network structure does not influence the simulation results, for this reason and considering that the SF network is the one that better reproduce the structure of online social media, we restrict our attention to SF networks.} Each agent holds an opinion $x_i$, $i\in\{1,\ldots, N\}$, uniformly distributed in $[0,1]$. Two agents interact if and only if they are connected in $G$ and  their present opinions are close enough, i.e. iff $j\in N_G(i)$ and $|x_i -x_j|<\varepsilon$, for $\varepsilon\in[0,1]$.\footnote{We apply periodic boundary conditions in the simulations and hence two users will interact if: $|x_i -x_j|_{\tau}<\varepsilon$, for $\varepsilon\in[0, 0.5]$.} If these conditions hold, the two agents change their opinions according to rule (\ref{bcm}), otherwise they do not interact at all:
	\begin{equation}\label{bcm}
	\begin{cases} x_i  =  x_i + \mu(x_j - x_i) \\ x_j  =  x_j + \mu(x_i - x_j) 
	\end{cases},
	\end{equation}
	where $\mu$ is taken in the interval $(0, 0.5)$.
	
It is known from previous studies \cite{ben2003bifurcations, ben2000multiscaling} that for $\varepsilon$ big enough consensus is reached.
The time rate change of  $\mathbb{P}(x, t)dx$, the fraction of agents whose opinion at time $t$ lies in the interval $[x, x+dx]$, is given by:
\begin{equation*}
\frac{\partial\mathbb{P}(x,t)}{\partial t} = - \mathbb{P}(x,t)\int_{-\varepsilon}^{\varepsilon} \mathbb{P}(x +y, t) dy
	+ \frac{1}{(1 - \mu)}\int_{-\varepsilon-2x}^{\varepsilon-2x} \mathbb{P}(x + y, t)\mathbb{P}(x-\frac{ \mu }{1 - \mu}y, t)dy.
\end{equation*}
The first two moments are given by $M_0 = \int \mathbb{P}(x,t)dx = 1$ and $ M_1 =\int x\mathbb{P}(x,t) dx = 0$, i.e. the total mass and the mean opinion,  are conserved \cite{ben2003bifurcations}. 
Let $\mathbb{P}(x,0) = 1$ be a flat initial condition, with $ x \in [0, 1]$. We are interested in the final state of the system $\mathbb{P}(x,\infty)$.
	
When all agents interact, i.e., when $\varepsilon \geq 1$ the rate equation is integrable.\footnote{As we use PBC, $\varepsilon \geq 1/2$ in the simulations.} The second moment obeys $\dot{M}_2 + M_0M_2/2= M_1^2$, and using $M_1 = 0 $ and $M_0 = 1$ we find that $M_2(t) = M_2(0)e^{-t/2}$, hence the second moment vanishes exponentially in time, all agents approach the center opinion and the system eventually reaches consensus \cite{ben2003bifurcations}:
$$
\mathbb{P}_{\infty}(x) = M_0\delta(x).
$$
When $\varepsilon\geq1$ the final state is a single peak located in the middle and, as long as $\varepsilon \geq 1/2$, this situation persists.\footnote{Again, thanks to the PBC we get $\varepsilon \geq 1/4$ in the simulations.} For smaller values of the threshold $\varepsilon$, it has been shown, by numerical simulations, that consensus  is not reached and the opinion evolves into clusters that are separated by a distance larger than $\varepsilon$. Once each cluster is isolated it evolves into a Dirac delta function as in the case $\varepsilon \geq 1$. 
The final distribution consists of a series of non interacting clusters at locations $x_i$ with masses $m_i$:
$$
\mathbb{P}_{\infty}(x) = \sum_{i = 1}^r m_i\delta(x - x_i),
$$
where $r$ is the number of evolving opinion	clusters  \cite{ben2003bifurcations}.
All clusters must fulfill the conservation laws $\sum m_i = M_0 = 1$, and $\sum x_im_i = M_1 = 0$ is equal to the conserved mean opinion. All different clusters $i \neq j$ must also fulfill $|x_i - x_j | > \varepsilon$.	

\section*{Results and Discussion}
\subsection*{Models}
Starting from the BCM we introduce three new models of opinion dynamics and network evolution. The first model we consider is the \textbf{Rewire with Bounded Confidence Model} (RBCM) that considers the same framework as in BCM and involves two phases. In phase one we run the \textit{rewiring steps} in which each agent $i$ interacts with a randomly chosen neighbor $j$ and, if the distance between the two opinions is above the tolerance $\varepsilon$, i.e. if $| x_i - x_j|_{\tau}  \geq \varepsilon$, for $\varepsilon\in[0, 0.5]$,\footnote{We restrict our attention to $\varepsilon\in[0, 0.5]$ after noticing that $\forall x, y \in\{1, \ldots, N\}$ we get $| x - y |_{\tau}\in[0, 0.5]$. We will assume $\varepsilon\in[0, 0.5]$ throughout the paper.} then their link is broken and $i$ is rewired to a randomly chosen agent $k\in\{1,\ldots,N\}/(N_G(i)\cup\{i\})$. Phase one ends when all links have an opinion distance below the tolerance $\varepsilon$. In phase two we run the BCM on the rewired network. As all the couples in the new network are concordant, all the randomly chosen pairs will interact and readjust their opinion according to the rule (\ref{bcm}).

The \textbf{Unbounded Confidence Model} (UCM) is the second of the models that we introduce and its novelty is to allow the interaction for every randomly chosen pair of neighbors $(i,j)$. To be specific, if two agents have concordant opinions, i.e. if $|x_i - x_j|_{\tau} < \varepsilon$, as for the previous model, we adjust $x_i$ and $x_j$ by rule (\ref{bcm}). However, if their opinions are discordant, i.e. if $|x_i - x_j|_{\tau} \geq \varepsilon$, we use a new updating rule (\ref{ucm}) that enables us to replicate the empirically observed repulsion of discordant opinions:
\begin{equation}\label{ucm}
\begin{cases}
x_i = x_i - \mu[x_j - x_i - \rho(x_j -x_i)]\\
x_j = x_j - \mu[x_i - x_j - \rho(x_i -x_j)]
\end{cases},
\end{equation}
where $\mu$ is taken in the interval $(0, 0.5)$ and $\rho(.)$ is defined in (\ref{pbc}). The adjustment $\rho(.)$ ensures the PBC by maintaining the opinions inside the interval $[0,1]$.		

The last model that we introduce is the\textbf{ Rewire with Unbounded Confidence Model} (RUCM) that again allows the interaction for every randomly chosen pair of users $(i,j)$ but at the same time allows for the random rewiring of discordant pairs. Specifically, if $|x_i - x_j|_{\tau} < \varepsilon$, then we adjust $x_i$ and $x_j$ by rule (\ref{bcm}). If $|x_i - x_j|_{\tau} \geq \varepsilon$, then we adjust $x_i$ and $x_j$ by rule (\ref{ucm}), the link between nodes $i$ and $j$ is broken, and a new link between $i$ and a randomly chosen user $k\in \{1, \ldots, N\}\backslash (N_G(i)\cup\{i, j\})$ is created.

\subsection*{Simulation Results}
\begin{figure}[h]
	\centering
	\includegraphics[scale=.8]{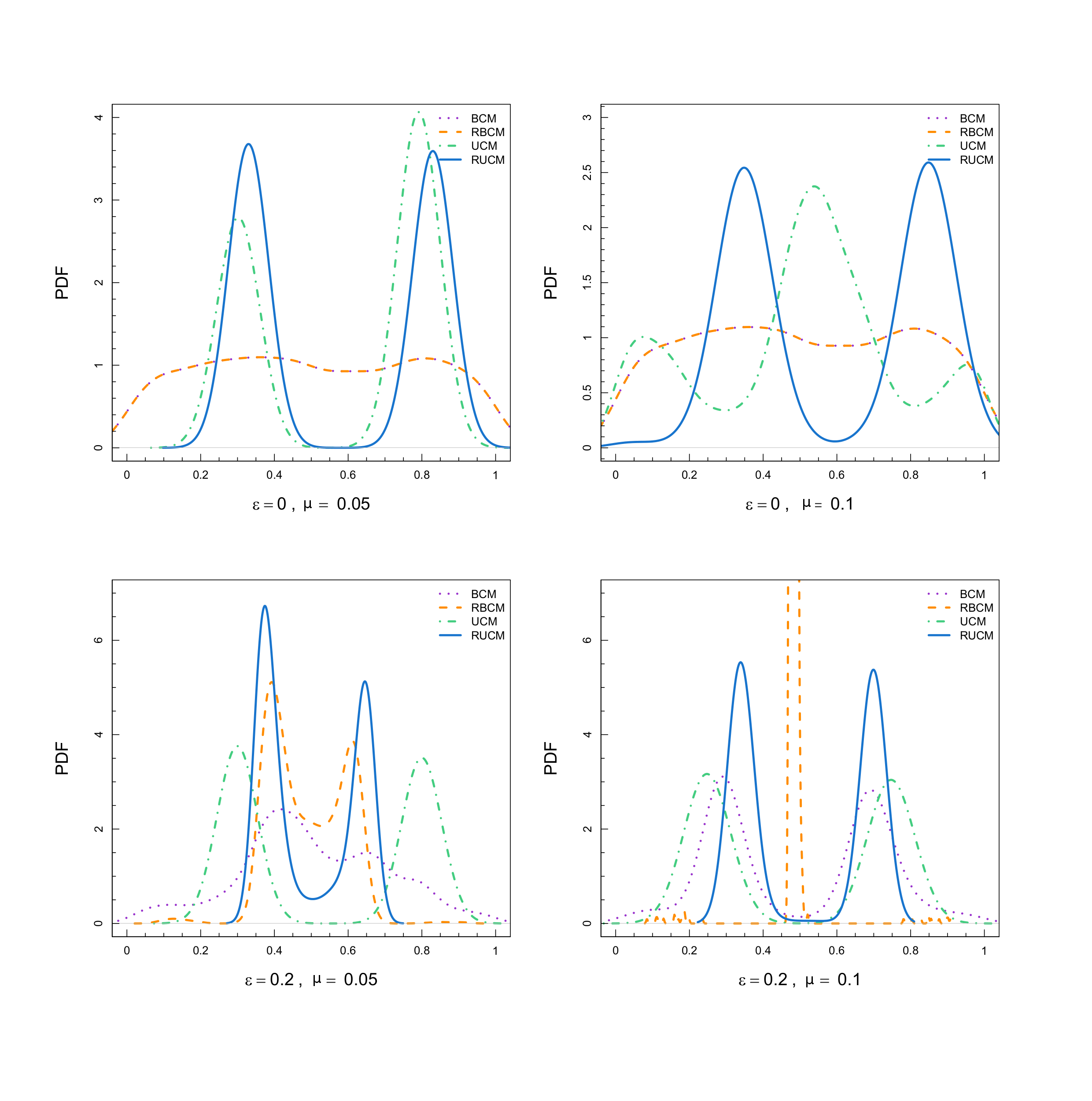}
	\caption{Probability density functions (PDFs) of final opinion, after a maximum of $10^5$ time steps or until convergence is reached, for four different combinations of the parameters $(\varepsilon,\mu)$. In the upper left figure we have $(\varepsilon,\mu) = (0,0.05)$, in the upper right $(\varepsilon,\mu) = (0,0.1)$, in the lower left $(\varepsilon,\mu) =  (0.2,0.05)$, and in the lower right $(\varepsilon,\mu) =  (0.2,0.1)$. In all figures the blue solid curve is for RUCM, the green dot-dashed one for UCM, the violet dotted one for BCM, and the pale orange dashed  one for RBCM. We observe a bimodal distribution for RUCM and UCM, representing the coexistence of two polarized stable opinions.}\label{density}
\end{figure}

\begin{figure}[h]
	\centering
	\includegraphics[scale=.3]{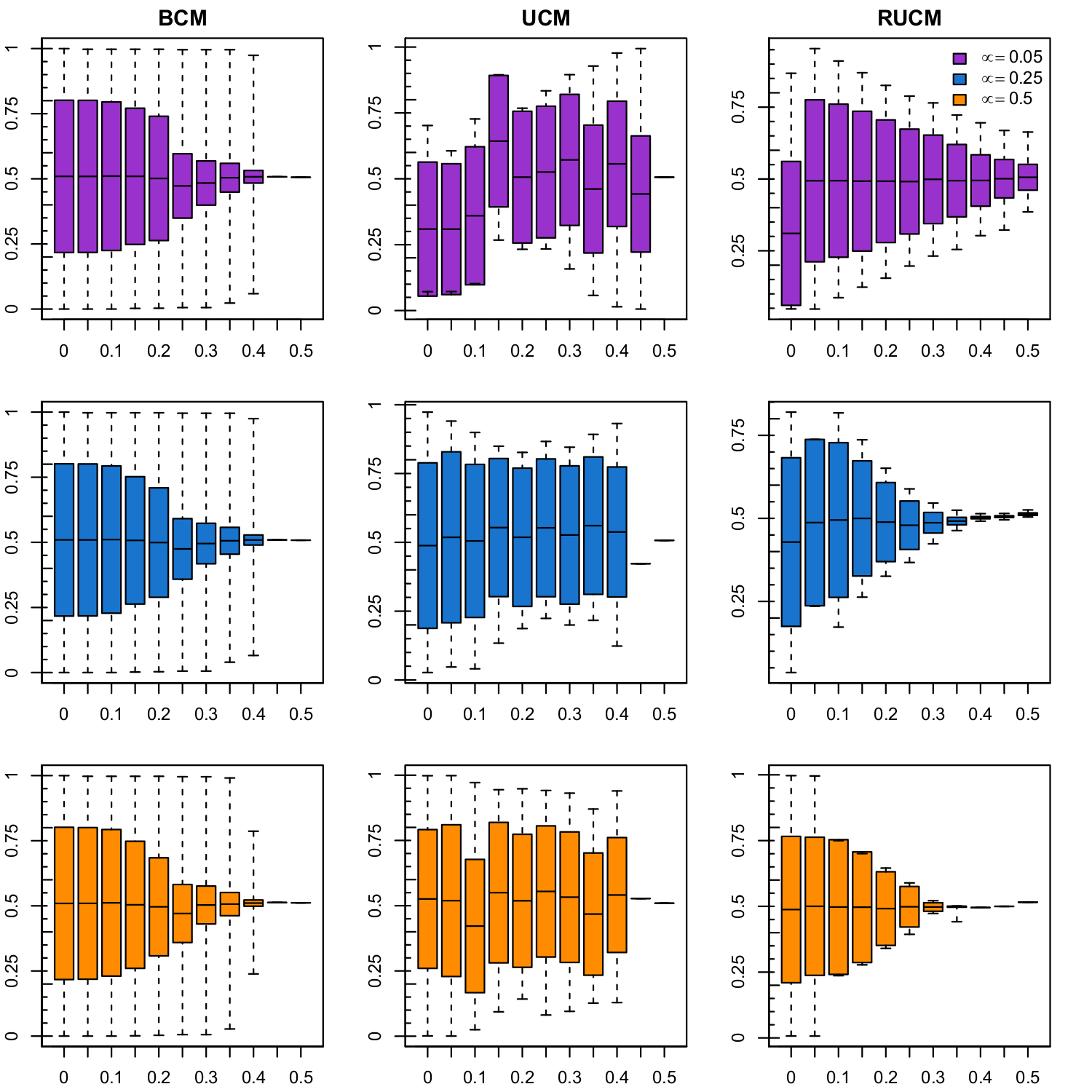}
	\caption{Summary statistics (mean, standard deviation, 1st quantile, and 3rd quantile) of the final opinion distributions for varying $\varepsilon$ and three  different values of $\mu$: violet denotes $\mu = 0.05$, blue denotes $\mu = 0.25$, and orange denotes $\mu = 0.5$. The left column is for BCM, the central one for UCM, and the right one for RUCM.}\label{summary}
\end{figure}

We perform Monte Carlo simulations of the BCM and the three new models on a scale-free network (SF) of 2000 nodes with the parameters $(\varepsilon, \mu)$ varying in the parameter space $[0, 0.5]\times[0, 0.5]$, for a maximum of $10^5$ times steps or until convergence is reached. We averaged our results over 5 repetitions. Figure \ref{density} shows the probability density functions (PDFs) of final opinion, after a maximum of $10^5$ time steps, for four different combinations of the pair of parameters $(\varepsilon,\mu)$: $(\varepsilon,\mu)\in\{(0, 0.05), (0, 0.1), (0.2, 0.05), (0.2, 0.1)\}$. The blue solid and the green dot-dashed curves refer to the newly introduced RUCM and UCM respectively, while the violet dotted curve is for BCM and the orange dashed for RBCM.
For all the parameter choices we observe a bimodal opinion distribution in the cases of RUCM and UCM (note that we assume periodic boundary conditions). It is interesting to note that for UCM and RUCM there are two polarized opinions also for $\varepsilon = 0$, while in that case BCM and RBCM show no changes with respect to the initial uniform distribution.
\clearpage
\clearpage
Figure \ref{summary} reports a collection of summary statistics (mean, standard deviation, 1st quantile, and 3rd quantile) of the final opinion distributions for varying $\varepsilon$ and three  different values of $\mu$ (violet is for $\mu = 0.05$, blue for $\mu = 0.25$, and orange for $\mu = 0.5$). The left column is for BCM, the central one for UCM, and the right one for RUCM. We omit the results for RBCM as we observe from the simulations that, after the rewiring steps, the dynamics are similar to the BCM case but with a faster convergence, refer to the \textit{Appendix} for an in depth analysis of the RBCM model.
We observe different mechanisms for the two newly introduced models, such as a faster convergence to the consensus state for RUCM. However, we need to study the final number of peaks to better characterize the differences between UCM and RUCM, and to relate them with the results for the classical BCM.

\subsection*{Final Distribution of Peaks}
We perform Monte Carlo simulations of the BCM, UCM, and RUCM on a scale-free network of 2000 nodes with $(\varepsilon, \mu )\in[0, 0.5]\times[0, 0.5]$, for a maximum of $10^5$ steps or until convergence is reached (the results are averaged over 5 repetitions). Given the final distributions of opinions obtained by the simulations, we compute the number of peaks of opinions as the local maxima in the distribution of frequencies of opinions. To be specific, we divide the interval $[0,1]$ in $100$ bins of length $0.01$ and consider the frequencies of values falling in each interval. We regard two peaks to be separate if the distance between the middle points of the respective bins is smaller than $0.1$. All the results are averaged over 5 repetitions.

\begin{figure}[h]
	\centering
	\includegraphics[scale=0.4]{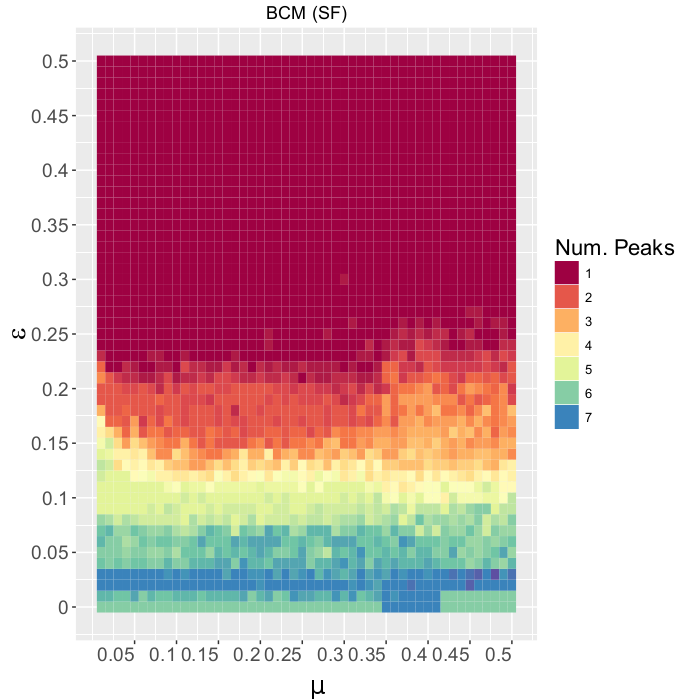}
	\caption{Final distribution of peaks for the BCM, with varying $(\varepsilon, \mu )\in[0, 0.5]\times[0, 0.5]$. The Monte Carlo simulations are carried on a Scale-Free network with 2000 nodes for a maximum of $10^5$ steps or until convergence is reached (all results are averaged over 5 repetitions).}\label{bcm_peaks}
\end{figure}

\begin{figure}[h]
	\centering
	\includegraphics[scale=1.3]{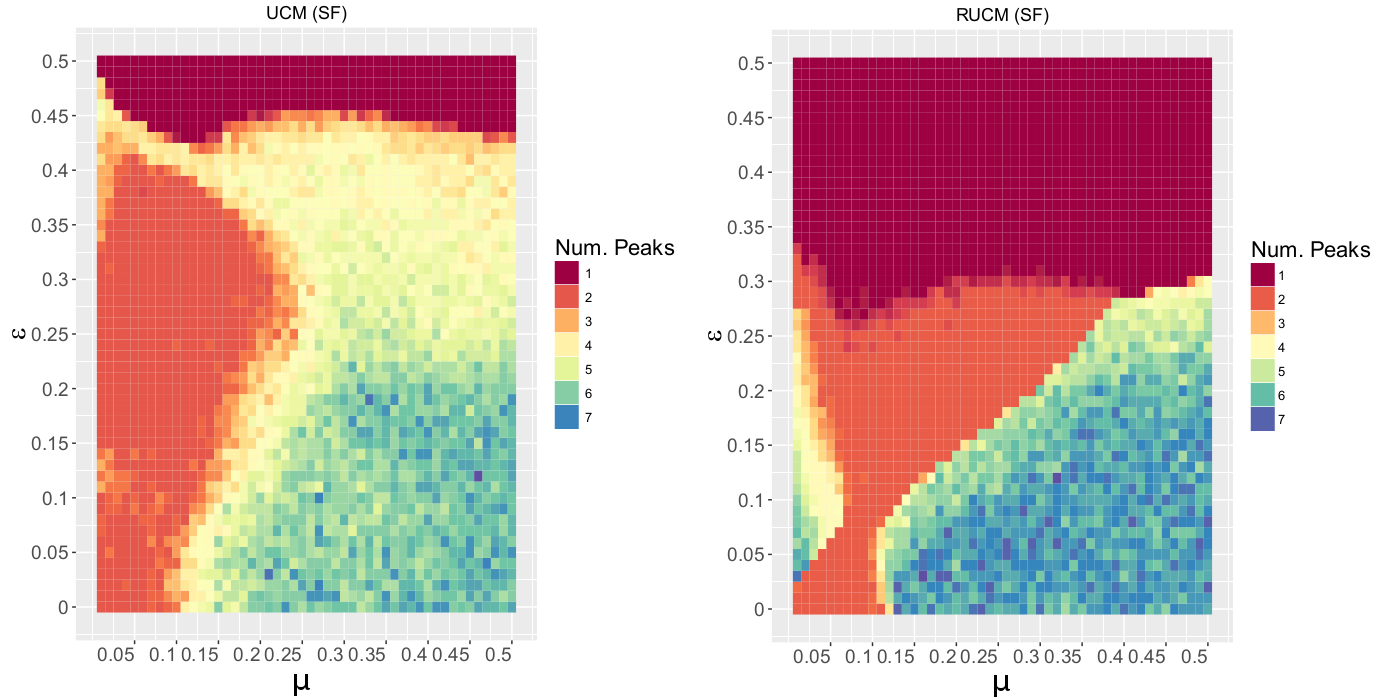}
	\caption{Final distribution of peaks for the UCM (left) and RUCM (right), with varying $(\varepsilon, \mu )\in[0, 0.5]\times[0, 0.5]$. The Monte Carlo simulations are carried on a Scale-Free network with 2000 nodes for a maximum of $10^5$ steps or until convergence is reached (all results are averaged over 5 repetitions).}\label{peaks}
\end{figure}

Figure \ref{bcm_peaks} shows the final distribution of peaks of BCM for varying  $(\varepsilon, \mu )\in[0, 0.5]\times[0, 0.5]$.\footnote{The corresponding result for the RBCM model is shown in Fig.\ref{app_peaks} of the \textit{Appendix}.} The final peaks distribution complies with theoretical \cite{ben2003bifurcations, ben2000multiscaling} and simulation \cite{deffuant2000mixing} results from previous work.
Figure \ref{peaks} shows the final peaks distribution of UCM (left) and RUCM (right) for varying $(\varepsilon, \mu )\in[0, 0.5]\times[0, 0.5]$. For both models we observe a large area of the parameter space for which two final opinions coexist.
We register a faster convergence to the consensus state for the RUCM (w.r.t UCM), that is due to the rewiring rule. Also, we observe that for the RUCM there is a direct transition from many opinions to two opinions, as well as from two opinions to consensus, while for the UCM there is an intermediate area where 3 or 4 opinions emerge, respectively shown in yellow and pale orange.

Comparing Figs. \ref{bcm_peaks} and \ref{peaks}, we see that the new models, unlike the BCM, are able to explain the coexistence of two stable final opinions, often observed in reality. Another important difference with respect to the BCM is that the $\mu$ parameter assumes an important role in tuning the number of final opinions peaks. The dependence of the number of final peaks on the $\mu$ parameter is stronger for the RUCM, where we observe a clear transition from many opinions to exactly two on the diagonal.

\subsection*{Mean Field Approximation}
For the RBCM, after the rewiring steps, all connected agents have an opinion distance below $\varepsilon$, meaning that they will always interact. The time rate of change of $\mathbb{P}(x,t)$ is equal to:

\begin{equation*}
\frac{\partial\mathbb{P}(x,t)}{\partial t}= - \mathbb{P}(x,t)\int_{-1}^{1} \mathbb{P}(x +y, t) dy + \frac{1}{(1 - \mu)}\int_{-1-2x}^{1-2x} \mathbb{P}(x + y, t)\mathbb{P}(x-\frac{ \mu }{1 - \mu}y, t)dy.
\end{equation*}
Considerations analogous to the BCM case hold (see the Section \textit{Material and Methods}). A faster convergence scale is also observed in the simulations.

In the UCM and RUCM case we consider two updating rules: rule (\ref{bcm}) if the opinions $(x_i,x_j)$ of the agents are close enough ($|x_i -x_j|_{\tau} < \varepsilon$) and rule (\ref{ucm}) if they are not ($|x_i -x_j|_{\tau} \geq \varepsilon$),. Thus the opinions will change according to $(x_i,x_j)\rightarrow(\hat{x}_i,\hat{x}_j)$:
$$
\bordermatrix{
	&        \cr
	& \hat{x}_i    \cr
	& \hat{x}_j    \cr
} =\bordermatrix {
&        \cr
& 1 - \vartheta_{\varepsilon}\mu + (1 - \vartheta_{\varepsilon})\mu & \vartheta_{\varepsilon}\mu - (1 - \vartheta_{\varepsilon})\mu   \cr
& \vartheta_{\varepsilon}\mu - (1 - \vartheta_{\varepsilon})\mu  & 1- \vartheta_{\varepsilon}\mu + (1 - \vartheta_{\varepsilon})\mu  \cr
} \bordermatrix {
&        \cr
& x_i    \cr
& x_j    \cr
}
+(1 - \vartheta_{\varepsilon})\mu \bordermatrix {
	&        \cr
	&\varrho(x_j - x_i) \cr
	& \varrho(x_i - x_j) \cr
} ,
$$
where $\vartheta_{\varepsilon} = \vartheta(\varepsilon - |x_i - x_j|_{\tau}))$ is the Heaviside theta function that equals $1$ if $\varepsilon - |x_i - x_j|_{\tau}<0$, $0$ otherwise, and $\varrho$ is defined in (\ref{pbc}).
There are two ways in which the density of opinion $x$ changes at every time step $t$: either an agent moves away from $x$ after an interaction ($I^-$) or she arrives in $x$ after an interaction ($I^+$). Let  $\mathbb{P}(x, t)dx$ be the fraction of agents whose opinion at time $t$ lies in the interval $[x, x+ dx]$, then its time rate of change is:
$$
\frac{\partial\mathbb{P}(x,t)}{\partial t} = I^{-}(x,t) + I^{+}(x,t).
$$
The negative part is defined as in the BCM case but for a wider interval:
$$
I^{-}(x,t) = - \mathbb{P}(x,t)\int_{-1}^{1} \mathbb{P}(x +y, t) dy,
$$
as $I^{-}(x,t)$ is simply the probability that an agent with opinion $x$ interacts with some other agent and thus moves away from $x$.
For $I^{+}(x,t)$ we have two terms depending on the distance of the initial opinions:
$$
I^{+}(x,t) = I^{+}_1(x,t) + I^{+}_2(x,t), 
$$
for the first term we get the same expression as in the BCM case:
$$
I^{+}_1(x,t) =  \frac{1}{(1 - \mu)}\int_{-\varepsilon-2x}^{\varepsilon-2x} \mathbb{P}(x + y, t)\mathbb{P}(x-\frac{ \mu }{1 - \mu}y, t)dy.
$$
For $I^{+}_2(x,t)$ we have to consider the negative update in (\ref{ucm}), and the integrals are over the interval for which $|x_1 - x_2|_{\tau} \geq \varepsilon$:
\begin{eqnarray*}
	I^{+}_2(x,t) &=& \int \int \mathbb{P}(x_1, t) \mathbb{P}(x_2, t) \delta(x+ \mu x_1 - (1 + \mu)x_2 -\mu\varrho))dx_1dx_2\\
	&=& \frac{1}{(1 + \mu)}\int  dx_1\mathbb{P}(x_1, t)\int\mathbb{P}(x_2, t)  \delta(x_2 -\frac{x + \mu x_1 -\mu\varrho}{1 + \mu} )dx_2\\
	&=& \frac{1}{(1 + \mu)}\int_{|x_1-x_2|_{\tau}\geq\varepsilon}\mathbb{P}(x_1, t) \mathbb{P}\left( \frac{x +  \mu(y- \varrho)}{1 + \mu}\right)dx_1\\
	&=& \frac{1}{(1 + \mu)}\int_{[-1,-\varepsilon - 2x]\cup [\varepsilon -2x, 1]}\mathbb{P}(x + y, t) \mathbb{P}\left(x + \frac{ \mu}{1 + \mu}(y- \varrho)\right)dy,
\end{eqnarray*}	
where $\varrho = \varrho_{x_2-x_1}$.
Hence we obtain:
\begin{eqnarray*}
	\frac{\partial\mathbb{P}(x,t)}{\partial t}  &=& - \mathbb{P}(x,t)\int_{-1}^{1} \mathbb{P}(x +y, t) dy + \frac{1}{(1 - \mu)}\int_{-\varepsilon-2x}^{\varepsilon-2x} \mathbb{P}(x + y, t)\mathbb{P}(x-\frac{ \mu }{1 - \mu}y, t)dy\\		
	&+& \frac{1}{(1 + \mu)}\int_{[-1,-\varepsilon - 2x]\cup [\varepsilon -2x, 1]}\mathbb{P}(x + y, t)\mathbb{P}\left(x + \frac{ \mu}{1 + \mu}(y-\varrho)\right)dy.
\end{eqnarray*}	
When all agents interact positively, i.e. when $\varepsilon \geq 1/2$, the third term of the rate equation disappears and we are again in the BCM case, where consensus is reached asymptotically and:
$$
\mathbb{P}_{\infty}(x) = M_0\delta(x). 
$$
From simulations results, we notice that the final state is a single peak as long as $\varepsilon\in(0.45,0.5)$ for the UCM, or $\varepsilon\in(0.3,0.5)$ for the RUCM (with the exception of those points for which $\mu$ is near to zero).

Unlike for BCM, in the new models the parameter $\mu$ plays an important role in the evolution of the distribution of opinions. For both UCM and RUCM we have the coexistence of two opinions in the final state for a wide region of the $(\varepsilon, \mu)$-plane, this region varies for the two models, in particular the faster convergence to the consensus state for the RUCM is due to the rewiring rule.
For smaller values of $\varepsilon$, and outside the two opinions region, we showed by numerical simulations that consensus is not reached, and many opinions at distance larger than $\varepsilon$ coexist.
	
\section*{Conclusions}
In recent years opinion dynamics has attracted much interest from the fields of  both statistical physics and social science. In classical models such  as the Sznajd model, the voter model, the majority rule model, and the bounded confidence model, consensus is eventually reached, under the correct conditions. However, in face-to-face and online opinion exchanges, consensus is not commonly achieved, and classical models fail to explain this empirically observed fact.

We propose a model of opinion dynamics capable of reproducing the empirically observed coexistence of two stable opinions. We assume the basic updating rule of the BCM and we develop two variations of the model: the \textit{Rewire with Bounded Confidence Model} (RBCM), in which discordant links are broken until convergence is reached; and the \textit{Unbounded Confidence Model}, under which the interaction among discordant pairs of users is allowed and a negative updating rule is introduced, either with the rewiring step (RUCM) or without it (UCM). 

From numerical simulations we find that the new models (UCM and RUCM), unlike the BCM, are able to explain the coexistence of two stable final opinions, often observed in reality. Another important difference with respect to the BCM is that the convergence parameter $\mu$ assumes an important role in tuning the number of final opinions peaks; hence, in our model the speed at which opinions converge/diverge allows to change the final opinion landscape. Lastly, we derive a mean field approximation of all the three new models.

\section*{Acknowledgements}
Funding for this work was provided by EU FET project MULTIPLEX nr. 317532, SIMPOL nr. 610704, DOLFINS nr. 640772, SOBIGDATA nr. 654024. The funders had no role in study design, data collection and analysis, decision to publish, or preparation of the manuscript.

\clearpage
\section*{Appendix}
\subsection*{Simulation Results for RBCM}

\begin{figure}[h]
	\centering
	\includegraphics[scale=0.25]{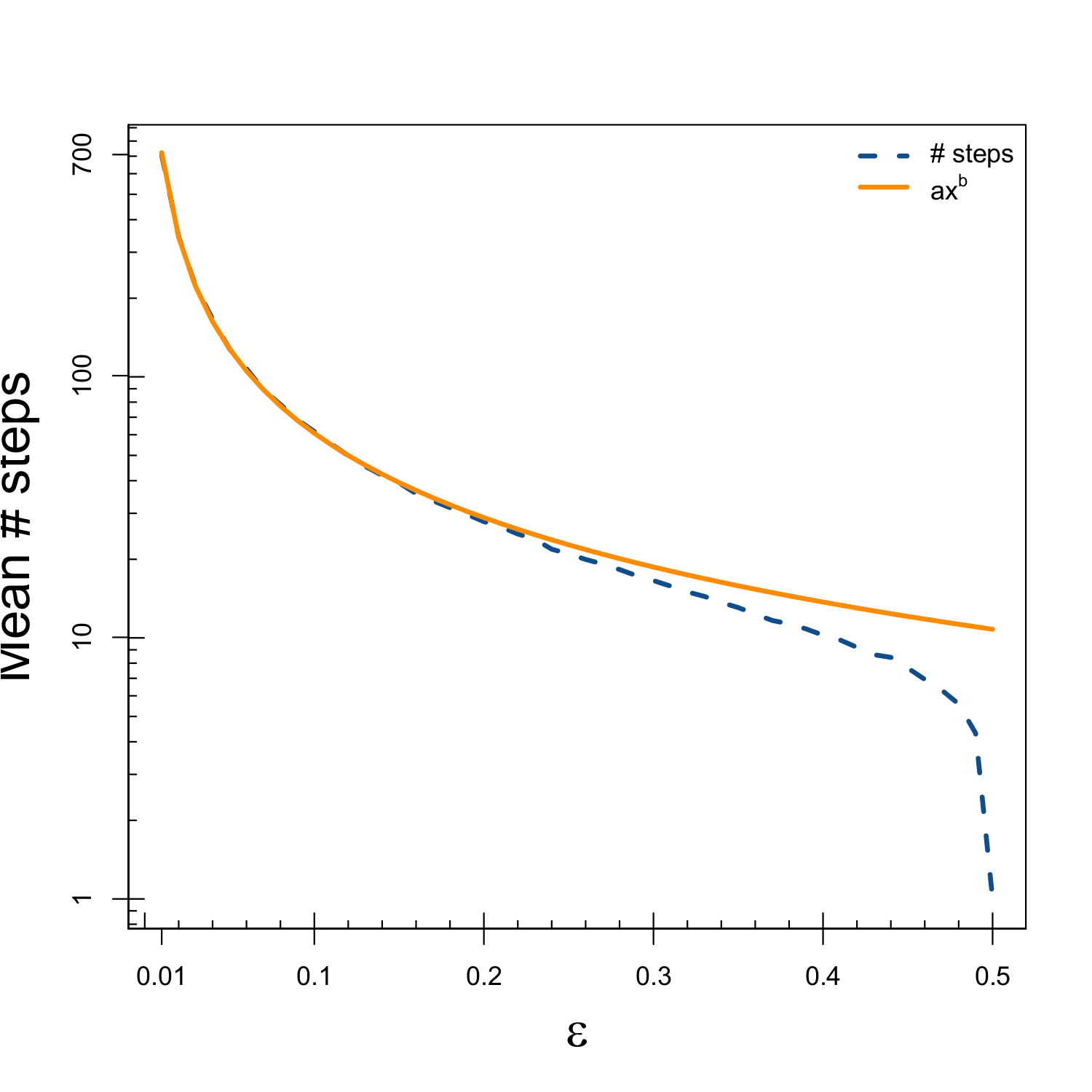}
	\caption{Estimated mean number of steps needed to get the fully concordant network as a function of $\varepsilon$ (dashed blue curve). The results are averaged over $50$ realizations. The decay of the estimated mean number of steps is best fitted by the power law $ax^{-b}$ (solid orange curve), where the parameters $a = 5.105$ and $b = 1.072$ are obtained through (NLS) fitting.}\label{app_steps}
\end{figure}

The RBCM differs from the standard BCM in the first phase, where a series of random rewiring steps is performed until all links in the network are concordant, meaning that the difference between the opinions of the two endpoints of each link is smaller than $\varepsilon$. Through this procedure we obtain a network in which all randomly chosen pairs of users interact and hence the consensus is reached also for smaller values of $\varepsilon$. In Fig. \ref{app_steps} we show the estimated mean number of steps needed to get the fully concordant network as a function of $\varepsilon$, where the results are averaged over $50$ realizations. The decay of the estimated mean number of steps is best fitted by the power law $ax^{-b}$, where the parameters $a = 5.105$ and $b = 1.072$ are obtained through Nonlinear Least Square (NLS) fitting.

\begin{figure}[h]
	\centering
	\includegraphics[scale=0.25]{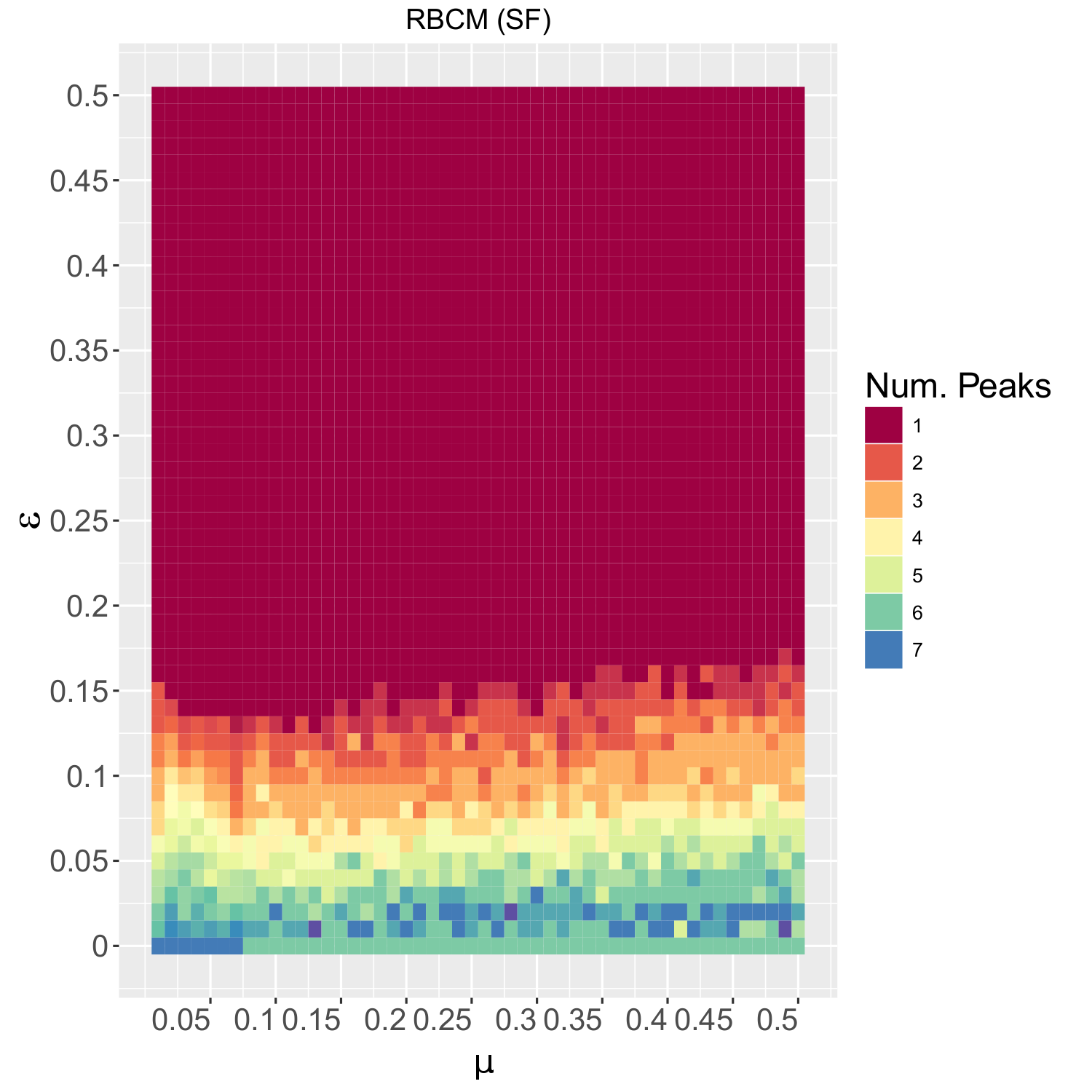}
	\caption{Final  distribution of peaks for the RBCM, with varying $(\varepsilon, \mu )\in[0, 0.5]\times[0, 0.5]$. The Monte Carlo simulations are carried on a Scale-Free network with 2000 nodes for a maximum of $10^5$ steps or until convergence is reached (the results are averaged over 5 repetitions).}\label{app_peaks}
\end{figure}

Figure \ref{app_peaks} shows the final distribution of peaks for the RBCM for varying  $(\varepsilon, \mu )\in[0, 0.5]\times[0, 0.5]$. We notice that consensus is reached for smaller values of $\varepsilon$ w.r.t the BCM. Indeed, while for BCM consensus is reached for $\varepsilon \geq 0.25$, for RBCM we get it for $\varepsilon\geq 0.15$.

\end{document}